\documentstyle[11pt,epsf,aaspp4]{article}

\def\n0LC{{n^{\rm LC}_0}}

\def\bft{{\vec\gamma}}
\def\nLC{{n^{\rm LC}}}

\def\calF{{\cal F}}

\def\n0LC{{n^{\rm LC}_0}}

\def\pp{\par\parshape 2 0truecm 15.5truecm 1truecm 14.5truecm\noindent}
\def\and{{$\&$ }}
\def\sigmav{{\sigma_v}}
\newcommand{\simgt}{\lower.5ex\hbox{$\; \buildrel > \over \sim \;$}}
\newcommand{\simlt}{\lower.5ex\hbox{$\; \buildrel < \over \sim \;$}}

\begin{document}

\begin{minipage}[c]{4cm}
HUPD-9907\\
RESCEU-16/99\\
UTAP-328/99\\
\end{minipage}\\

\title{The cosmological light-cone effect on the power spectrum \\
of galaxies and quasars in wide-field redshift surveys}

\bigskip

\author{Kazuhiro Yamamoto, Hiroaki Nishioka 
\\ {\it Department of Physics, Hiroshima
    University, Higashi-Hiroshima 739-8526, Japan.} \\ and \\ 
Yasushi Suto \\ {\it Department of Physics and Research Center for
    the Early Universe (RESCEU) \\ School of Science, University of
    Tokyo, Tokyo 113-0033, Japan.} }

\received{1999 May 12}
\accepted{1999 July ??}

\bigskip

\begin{abstract}
  We examine observational consequences of the cosmological light-cone
  effect on the power spectrum of the distribution of galaxies and
  quasars from upcoming redshift surveys. First we derive an
  expression for the power spectrum of cosmological objects in real
  space on a light cone, $P^{\rm LC}_{\rm R,lin}(k)$, which is exact
  in linear theory of density perturbations. Next we incorporate
  corrections for the nonlinear density evolution and redshift-space
  distortion in the formula in a phenomenological manner which is
  consistent with recent numerical simulations. On the basis of this
  formula, we predict the power spectrum of galaxies and quasars on
  the light cone for future redshift surveys taking account of the
  selection function properly.  We demonstrate that this formula
  provides a reliable and useful method to compute the power spectrum
  on the light cone {\it given an evolution model of bias}.
\end{abstract}


\newpage
\section{Introduction}

The importance of the cosmological light-cone effect in the future
redshift surveys, including the Two-degree Field (2dF) and the Sloan
Digital Sky Survey (SDSS), has been well recognized recently (e.g.,
Matarrese et~al. 1997; Matsubara, Suto, \& Szapudi 1997; Nakamura,
Matsubara, \& Suto 1998; Moscardini et al. 1998; de Laix \& Starkman
1998). In a series of previous papers (Yamamoto \& Suto 1999, Paper I;
Nishioka \& Yamamoto 1999a, Paper II; Suto et~al.  1999, Paper III),
we have extensively explored and discussed various aspects of the
cosmological light-cone effect on the two-point correlation functions,
$\xi^{\rm LC}(r)$, of objects at high redshifts.  Extending the
procedure described in these papers, we develop here a formula to
predict the power spectrum of cosmological objects located on the
light cone, $P^{\rm LC}(k)$.

Inhomogeneities of the matter distribution might affect the light
propagation, i.e., the gravitational lensing effect. In fact, the
angular correlation functions are significantly affected by the
gravitational lensing through the magnification bias. The spatial
correlations which we study in the present paper, however, are fairly
insensitive to the effect (e.g., Moessner, Jain, \& Villumsen 1998).
Therefore we ignore lensing effect below.

The plan of the paper is as follows. In \S 2, we define an estimator
for the power spectrum constructed from a sample of cosmological
sources on a light cone, which generalizes the conventional power
spectrum defined on the constant-time hypersurface. By computing the
ensemble average of the estimator, we derive the power spectrum on the
light cone. A mathematical derivation is outlined in Appendix A.
While this rigorous procedure assumes linear theory of density
perturbations in real space, the effects of nonlinear gravitational
growth and redshift-space distortion are important in practice (Papers
II and III). Thus we incorporate those effects as described in \S 3.
Then we apply our theoretical results to galaxy and quasar samples
with the SDSS spectroscopic samples specifically in mind (\S 4).
Finally \S 5 is devoted to discussion and conclusions.  Throughout
this paper we use the units in which the light velocity $c$ is unity.

\section{Power spectrum on the light cone: linear theory in real space}

\subsection{Basic procedure}

In this section we describe our theoretical formulation for the power
spectrum on a light-cone hypersurface which proceeds fairly parallel
to Paper I.  For simplicity, we focus on the spatially-flat
Friedmann-Lema\^{i}tre universe, whose line element is expressed in
terms of the conformal time $\eta$ as
\begin{equation}
  ds^2 = a^2(\eta) \left[-d\eta^2+d\chi^2+\chi^2 d\Omega_{(2)}^2 \right] .
\label{metric}
\end{equation}
We normalize the scale factor to be unity at present, $a(\eta_0)=1$.
All cosmological objects observed from redshift surveys are located on
the light-cone hypersurface of the space-time (\ref{metric}), defined
by an observer.  We locate the fiducial observer at the origin of the
coordinates ($\eta=\eta_0$, $\chi=0$), and therefore an object at
$\chi$ and $\eta$ on the light-cone hypersurface of the observer
satisfies a simple relation $\eta=\eta_0-\chi$. Then the (real-space)
position of the source on the light-cone hypersurface is specified by
$(\chi,\bft)$, where $\bft$ is a unit vector along the line-of-sight.
In order to avoid confusion, we introduce a radial coordinate $r$
instead of $\chi$, and write the metric of the three-dimensional real
space as
\begin{equation}
ds_{\rm LC}^2 = dr^2+r^2d\Omega_{(2)}^2 .
\label{LCmetric}
\end{equation}

Let us denote the number density field of the sources on the light
cone by $\nLC(r,\bft)$, which is simply related to the comoving number
density of objects at a conformal time $\eta$ and at a position
$(\chi,\bft)$ as
\begin{equation}
  \nLC(r,\bft)=n(\eta,\chi,\bft)~\bigr|_{
  \eta\rightarrow \eta_0-r, ~\chi\rightarrow r}~.
\label{eq:nlc1}
\end{equation}
Introducing the mean {\it observed} (comoving) number density
$n_0(\eta)$ at time $\eta$ and the density fluctuation of 
luminous objects $\Delta(\eta,\chi,\bft)$, we write
\begin{equation}
  n(\eta,\chi,\bft) = n_0(\eta) \left[1+\Delta(\eta,\chi,\bft)\right] .
\label{defDelta}
\end{equation}
Then equation (\ref{eq:nlc1}) is rewritten as 
\begin{equation}
  \nLC(r,\bft)=n_0(\eta) \left[1+\Delta(\eta,\chi,\bft) 
  \right]~\bigr|_{\eta\rightarrow \eta_0-r, ~\chi\rightarrow r} .
\label{nLC}
\end{equation}
In what follows, we define
\begin{equation}
  n^{\rm LC}_0(r) \equiv n_0(\eta)~\bigr|_{\eta\rightarrow \eta_0-r}~
\label{n0LC}
\end{equation}
for convenience.  Note that $n_0(\eta)$ is different from the mean
number density of the objects $\overline{n}(\eta)$ at $\eta$ by the
selection function which depends on the luminosity function of the
objects and thus on the magnitude-limit of a survey (see Paper I and
\S 4 below).

Following the conventional treatment of the power spectrum in
cosmology (e.g., Feldman, Kaiser, \& Peacock 1994), we introduce the
number density field of cosmological objects on the light cone:
\begin{equation}
  F(r,\bft)={n^{\rm LC}(r,\bft)- n_s^{\rm LC}(r,\bft) 
           \over 
\displaystyle \left[\int d^3{\bf r} \n0LC(r)^2\right]^{1 \over 2}},
\label{1}
\end{equation}
where $n_s^{\rm LC}(r,\bft)$ denotes the number density of the synthetic 
catalog without structure, which has the mean number density 
$\n0LC(r)$. To be specific, it satisfies
\begin{eqnarray}
  &&\bigl<n_s^{\rm LC}(r_1,\bft_1)n_s^{\rm LC}(r_2,\bft_2)\bigr>
  =\bigl<n^{\rm LC}(r_1,\bft_1)n_s^{\rm LC}(r_2,\bft_2)\bigr>
  =\n0LC(r_1) \n0LC (r_2) ,
\label{13}
\end{eqnarray}
where we neglect the Poisson noise term.

A power spectrum on the light cone, which can be estimated from a given
survey, is written as
\begin{equation}
  P(k)^{\rm obs}={\displaystyle 
\int_{V_s} d^3{\bf k} \vert \calF({\bf k }) \vert^2
        \over   \displaystyle\int_{V_s} d^3{\bf k}},
\label{eq:pkobs}
\end{equation}
where $V_s$ is the volume of a thin shell with the radius $k$ 
in the Fourier space, and $\calF({\bf k})$ is the Fourier
transform of equation (\ref{1}):
\begin{equation}
  \calF({\bf k})={\displaystyle\int d^3{\bf r}
  [n^{\rm LC}(r,\bft)-n^{\rm LC}_s(r,\bft)] e^{i\bf k\cdot r}
  \over \displaystyle \left[\int d^3{\bf r} \n0LC(r)^2\right]^{1 \over 2}}~.
\end{equation}

Finally the power spectrum on the light cone, $P^{\rm LC}_{\rm
  R,lin}(k)$, can be computed by taking the ensemble average of the
above estimator (\ref{eq:pkobs}):
\begin{eqnarray}
&&  P^{\rm LC}_{\rm  R,lin}(k) = \bigl<P(k)^{\rm obs}\bigr>
=\bigl< |\calF({\bf k})|^2 \bigr> \nonumber \\ 
&& = {\displaystyle\int d^3{\bf r}_1 \int d^3 {\bf r}_2 
  \bigl<[n^{\rm LC}(r_1,\bft_1)-n_s^{\rm LC}(r_1,\bft_1)]
  [n^{\rm LC}(r_2,\bft_2)-n_s^{\rm LC}(r_2,\bft_2)]\bigr>
  e^{i {\bf k} \cdot ({\bf r}_1-{\bf r}_2)}
  \over \displaystyle\int d^3{\bf r} n_0^{\rm LC}(r)^2} .
\label{eq:defPkLC}
\end{eqnarray}
For definiteness, we use a superscript LC explicitly to indicate the
power spectrum of {\it objects on the light cone} throughout the
paper. The power spectrum without the superscript denotes that of {\it
  mass} defined on the constant-time hypersurface.  Also subscripts
$R$ and $S$ indicate those in real and redshift spaces, and subscripts
lin and nl refer to those in linear and nonlinear models.

Assuming linear bias between density fluctuations of mass and luminous
objects and also linear theory for growth of density fluctuations, we
find that equation (\ref{eq:defPkLC}) reduces to the following
expression:
\begin{eqnarray}
  P^{\rm LC}_{\rm  R,lin}(k) &=& \left[\int dr r^2 \n0LC(r)^2\right]^{-1}
  \int dr_1r_1 \int dr_2 r_2 \n0LC(r_1)\n0LC(r_2)
  D_1(\eta_0-r_1)D_1(\eta_0-r_2)
\nonumber\\
  &~& \hspace{0.1cm}\times{1\over 2\pi k}\int^\infty_0dk_1k_1 
P_{\rm  R,lin}(k_1,\eta_0)
  b(k_1;\eta_0-r_1)b(k_1;\eta_0-r_2) \Pi(r_1,r_2,k_1,k)~,
\label{PLC}
\end{eqnarray}
where $D_1(\eta)$ is the linear growth rate normalized unity at
present, $D_1(\eta_0)=1$, $b(k;\eta)$ is the linear bias factor, 
and 
\begin{eqnarray}
\Pi(r_1,r_2,k_1,k) &\equiv& -C_i\bigl(|r_1-r_2||k_1-k|\bigr)
                         +C_i\bigl(|r_1-r_2|(k_1+k)\bigr)
\nonumber\\
  &~& \hspace{2cm}       +C_i\bigl((r_1+r_2)|k_1-k|\bigr)
                         -C_i\bigl((r_1+r_2)(k_1+k)\bigr)
\label{defPi}
\end{eqnarray}
with the cosine integral function $C_i(x)$ defined as
\begin{equation}
  C_i(x)=-\int^\infty_x dt{\cos t\over t}.
\end{equation} 
The derivation of equation (\ref{PLC}) from equation (\ref{eq:defPkLC})
is quite tedious but straightforward which we outline in Appendix A.

\subsection{Approximation to the exact formula}

While equation (\ref{PLC}) for power spectrum on the light cone is
exact as long as linear theory of density fluctuations and a linear
bias model are adopted, it looks fairly complicated.  Therefore it is
instructive to consider a practical approximation to equation
(\ref{PLC}). One can show that the function $\Pi(r_1,r_2,k_1,k)$
(eq.[\ref{defPi}]) is peaked around $r_1\simeq r_2$ and $k\simeq k_1$.
In fact, if one replaces the function as 
\begin{eqnarray}
  \Pi(r_1,r_2,k_1,k)\simeq 2\pi\delta(k_1-k)\delta(r_1-r_2)~ ,
\label{Deltaasym}
\end{eqnarray}
equation (\ref{PLC}) reduces to
\begin{eqnarray}
  P^{\rm LC}_{\rm  R,lin}(k)\simeq \alpha_{\rm  R,lin}(k) 
P_{\rm  R,lin}(k,z=0),
\label{PLCB}
\end{eqnarray}
with
\begin{equation}
  \alpha_{\rm  R,lin}(k)=\left[\int dr r^2 \n0LC(r)^2\right]^{-1}
    {\int dr r^2 n_0^{\rm LC}(r){}^2 
             D_1(\eta_0-r)^2} b(k;\eta_0-r)^2~.
\label{alphaD}
\end{equation}
In an extreme case of no evolution without biasing, i.e.,
$b(k;\eta)=D_1(\eta)=1$, $\alpha_{\rm R,lin}(k)$ becomes unity, and
$P^{\rm LC}_{\rm R,lin}(k)$ is equivalent to $P(k)_{\rm R,lin}$ as
expected. Thus if the approximation (\ref{Deltaasym}) is justified,
$\alpha_{\rm R,lin}(k)$ quantifies the degree of the light-cone effect
on a specified survey, which sensitively depends on the observed
selection function and the assumed time-evolution of linear bias.

\begin{figure}[t]
\centerline{\epsfxsize=12cm \epsffile{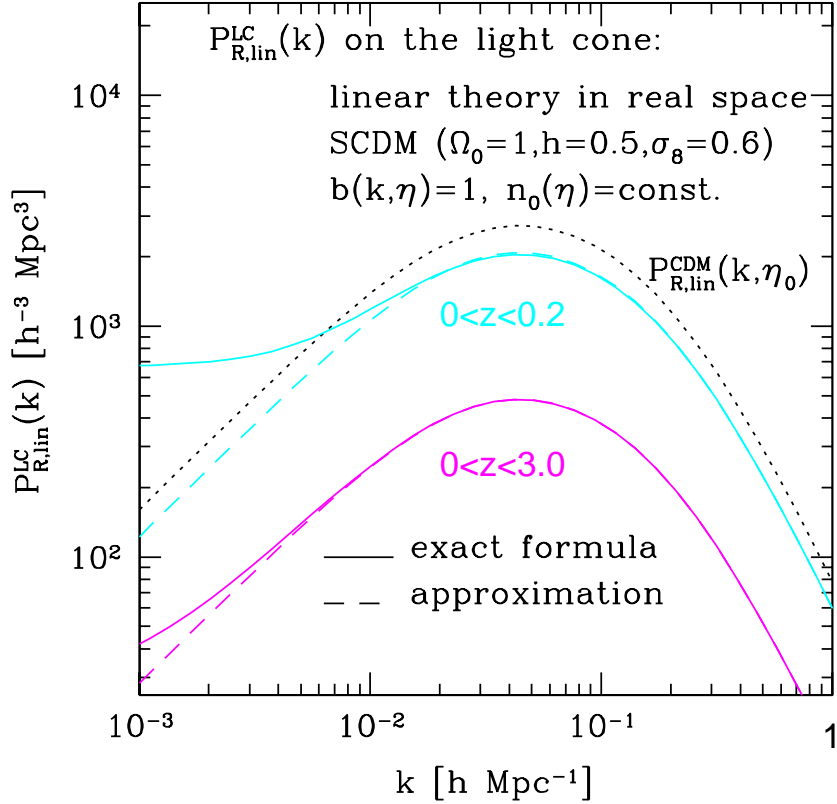}}
   \caption{
    Power spectra of cosmological objects on a light cone,
    $P^{\rm LC}_{\rm R,lin}(k)$.  Upper and lower pairs of curves
    correspond to the samples of survey depth extending up to $z=0.2$
    and $3.0$, respectively (the exact formula
    eq.[\protect{\ref{PLC}}] in solid lines, and its approximation
    eq.[\protect{\ref{PLCB}}] in dashed lines).  For comparison, the
    power spectrum of the mass fluctuations defined on the
    constant-time hypersurface at present is plotted in dotted line.
    For definiteness we adopt SCDM model in which $\Omega_0=1$,
    $\Omega_\Lambda=0$, $h=0.5$ and $\sigma_8=0.6$
}
    \label{fig:pklcapprox}
\end{figure}

Figure \ref{fig:pklcapprox} compares the power spectra on a light
cone, equations (\ref{PLC}) and (\ref{PLCB}). It should be noted
that $P^{\rm LC}_{\rm R,lin}(k)$ in the exact formula becomes
significantly larger than $P_{\rm R,lin}(k)$ for $k \simlt 1/r(z_{\rm
  max})$ with $r(z_{\rm max})$ being the depth of the survey volume.
This artifact is not particular to the light-cone effect, but simply
originates from the fact that the power on scales comparable to, or
larger than, the finite size of the survey volume cannot be properly
evaluated.

Thus except for the range $k \ll 1/r(z_{\rm max})$, where the estimate
of the power is fairly reliable for a given observational sample,
Figure \ref{fig:pklcapprox} shows that the approximation (\ref{PLCB})
reproduces the exact formula very accurately.  Therefore in what
follows, we will incorporate several important effects (redshift-space
distortion, nonlinear density and velocity evolution, and the
observational selection function) on the basis of the approximation.

While the results in Figure \ref{fig:pklcapprox} do not include the
selection function, i.e., $n_0(z)$=constant. is assumed, we also
examined the case with the proper selection functions for galaxies and
quasars (\S 4), and made sure that our approximation agrees with the
exact formula within several percents at worst.  This level of
accuracy is sufficiently good in the light of the other approximations
employed in describing the nonlinearity of density and velocity
fields, most notably, the uncertainty of the model of bias.

\section{Redshift-space distortion and nonlinear evolution of density
and velocity fields}

The results presented in the previous section are unrealistically
simplified since we have neglected several important effects including
the redshift-space distortion due to the peculiar velocity effect
(Kaiser 1987), nonlinear evolution of mass density fluctuations and of
peculiar velocity dispersions (finger-of-God), evolution of the bias
parameter (e.g., Fry 1996), and the observational selection function
which depends on a specific survey strategy.  In this section we
describe our modeling to the first three effects which are the
necessary ingredients in completing the theoretical predictions.  The
selection function is discussed in the next section.  Strictly
speaking, inclusion of these effects invalidates the exact derivation
of the formula (\ref{PLC}) unfortunately. Therefore we implement a
{\it phenomenological} correction to these effects as described below,
but we believe that this procedure is important and useful in
practice, especially given the uncertainty of the bias evolution
itself.

\subsection{Linear redshift-space distortion}

The cosmological observation is possible only in redshift space.  The
redshift-space distortion in linear theory, first discussed by Kaiser
(1987), is easily included in our formula for power spectrum on the
light cone in real space. In linear theory, the direction--averaged
power spectrum in redshift space, $P_{\rm S,lin}(k,z)$ is related to that in
real space, $P_{\rm R,lin}(k,z)$, as
\begin{equation}
  P_{\rm S,lin}(k,z)=\left[1+{2\over3}\beta(z)+{1\over5}\beta^2(z) \right] 
P_{\rm R,lin}(k,z) .
\label{defPS}
\end{equation}
The above relation (Kaiser 1987) assumes a distant-observer
approximation, and scale-independent linear bias, $b(z)$, as well. Then
$\beta(z)$ is defined by
\begin{equation}
  \beta(z)= {1\over b(z)}{d \ln D_1 \over d\ln a}.
\end{equation}
See Matsubara \& Suto (1996), Hamilton (1998) and Paper III for
extensive discussion on the redshift-space distortion.

Substituting equation (\ref{defPS}) in equation (\ref{PLC}) 
yields the expression for $P^{\rm LC}_{\rm S,lin}(k)$ which is approximated as
\begin{eqnarray}
 P^{\rm LC}_{\rm S,lin}(k) &\simeq& \alpha_{\rm S,lin}(k) P_{\rm R,lin}(k,z=0),
\label{eq:PLCSlin} \\
\alpha_{\rm S,lin}(k) &=& 
{\displaystyle {\int dr r^2 n_0^{\rm LC}(r)^2 D_1(\eta_0-r)^2} b(k;\eta_0-r)^2
\left[1+{2\over3}\beta(\eta_0-r)+{1\over5}\beta^2(\eta_0-r) \right] 
\over
{\displaystyle \int dr r^2 \n0LC(r)^2 }} .
\hspace*{0.5cm}
\label{eq:aSlin}
\end{eqnarray}
A similar expression for the linear redshift-space distortion on
two-point correlation functions is derived in Paper II.

\subsection{Nonlinear evolution of density}

Nonlinear evolution of mass density field introduces the additional
$k$-dependence on the light-cone power spectrum. While the full
description of the nonlinear gravitational evolution of density fields
is almost impossible, fairly accurate fitting formulae for
$P_{\rm R,nl}(k,z)$ have been obtained (e.g., Peacock \& Dodds 1994,1996).
We attempt to correct for the nonlinear density evolution simply by
adopting the fitting formulae, and then obtain an approximate expression
for $P^{\rm LC}_{\rm R,nl}(k)$ as follows:
\begin{eqnarray}
  P^{\rm LC}_{\rm R,nl}(k) &\simeq& \alpha_{\rm R,nl}(k) P_{\rm R,lin}(k,z=0),
\label{eq:PLCRnl} \\
  \alpha_{\rm R,nl}(k) &=& 
{\displaystyle {\int dr r^2 n_0^{\rm LC}(r)^2 } b(k;\eta_0-r)^2
P_{\rm R,nl}(k,\eta_0-r)
\over P_{\rm R,lin}(k,z=0) {\displaystyle \int dr r^2 \n0LC(r)^2} }.
\label{eq:aRnl}
\end{eqnarray}

\subsection{Nonlinear redshift-space distortion; finger-of-God effect}

Finally an effect of the nonlinear velocity field, {\it
  finger-of-God}, should be taken into account in order to make a
testable theoretical prediction. Again this has been extensively
discussed for the power spectrum on the constant-time hypersurface
(Cole, Fisher \& Weinberg 1994; Mo, Jing, \& B\"{o}rner 1997; Paper
III; Magira, Jing \& Suto 1999). In particular, Cole et al. (1994)
proposed the following correction for the finger-of-God effect:
\begin{equation}
  P_{\rm S,nl}(k,z)=
\left[A(\kappa)+{2\over3}\beta(z) B(\kappa)+{1\over 5}\beta^2(z) 
C(\kappa)\right]  P_{\rm R,nl}(k,z) .
\label{eq:PSnl}
\end{equation}
If the velocity distribution function is approximated by an
exponential model with scale-independent one-dimensional dispersion of
the peculiar velocity $\sigmav(z)$, which is indicated
observationally (Davis \& Peebles 1983) and also from numerical
simulations (Efstathiou et al. 1988; Ueda, Itoh \& Suto 1993; Magira
et al. 1999), the above functions are explicitly given by
\begin{eqnarray}
A(\kappa) &=& {{\rm arctan}(\kappa/\sqrt{2})\over \sqrt{2}\kappa}+
  {1\over 2+\kappa^2},
\\
  B(\kappa) &=& {6\over\kappa^2}\biggl(A(\kappa)-{2\over 2+\kappa^2}\biggr),
\\
  C(\kappa) &=& {-10\over\kappa^2}\biggl(B(\kappa)-{2\over 2+\kappa^2}\biggr),
\end{eqnarray}
with $\kappa(z)=k\sigmav(z)/H_0$. On large scales, $\sigmav(z)$ can be
well approximated by a fitting formula proposed by Mo, Jing \&
B\"{o}rner (1997). The validity and limitation of the approximation
are discussed in detail by Magira et al. (1999). Combining equation
(\ref{eq:PSnl}) with those fitting formulae, one can compute the
nonlinear power spectrum on the light cone in redshift space as
\begin{eqnarray}
  P^{\rm LC}_{\rm S,nl}(k) &\simeq& \alpha_{\rm S,nl}(k) P_{\rm R,lin}(k,z=0),
\label{eq:PLCSnl} \\
  \alpha_{\rm S,nl}(k) &=& 
{\displaystyle {\int dr r^2 n_0^{\rm LC}(r)^2 } b(k;\eta_0-r)^2
 P_{\rm R,nl}(k,\eta_0-r)\left[A(\kappa)+{2\over3}\beta B(\kappa)
+{1\over 5}\beta^2 C(\kappa)\right]
\over P_{\rm R,lin}(k,z=0) {\displaystyle \int dr r^2 \n0LC(r)^2} }.
\hspace*{1cm}
\label{eq:aSnl}
\end{eqnarray}

\section{Predictions of power spectra on the light cone 
for the future redshift surveys}

\subsection{Selection function for galaxies and quasars}

In properly predicting the power spectra on the light cone, the
selection function should be specified. In this subsection, we describe
the selection functions of galaxies and quasars with the upcoming SDSS
spectroscopic samples in mind.

For galaxies, we adopt a B-band luminosity function of the APM galaxies
(Loveday et~al. 1992) fitted to the Schechter function:
\begin{equation}
  \phi(L)dL=\phi^*\biggl({L\over L^*}\biggr)^\alpha
  \exp\biggl(-{L\over L^*}\biggr)d\biggl({L\over L^*}\biggr), 
\end{equation}
with $\phi^*=1.40\times 10^{-2} h^3{\rm Mpc}^{-3}$, $\alpha=-0.97$, and
$M_{B}^*=-19.50+5\log_{10} h$.  Then the comoving number density of
galaxies at $z$ which are brighter than the limiting magnitude $B_{\rm
lim}$ is given by
\begin{eqnarray}
  n_0(z,<B_{\rm lim})=\int_{L(B_{\rm lim},z)}^\infty \phi(L) dL
  =\phi^* \Gamma[(\alpha+1,x(B_{\rm lim},z)] ,
\end{eqnarray}
where 
\begin{equation}
  x(B_{\rm lim},z) \equiv {L(B_{\rm lim},z) \over L^*}
= \left[{d_L(z)\over 1 h^{-1}{\rm Mpc}}\right]^2 10^{2.2-0.4 B_{\rm lim}},
\end{equation}
and $\Gamma[\nu,x]$ is the incomplete Gamma function.

For quasars, we compute a selection function for B-band magnitude
limited samples following Paper I on the basis of the luminosity
function by Wallington \& Narayan (1993; see also Nakamura \& Suto
1997).

\subsection{Models and predictions}

For definiteness, we consider SCDM (standard cold dark matter) and
LCDM (Lambda cold dark matter) models, which have $(\Omega_0,
\Omega_\Lambda, h, \sigma_8)$ $= (1.0, 0.0, 0.5, 0.6)$ and $(0.3, 0.7,
0.7, 1.0)$, respectively. These sets of cosmological parameters are
chosen so as to reproduce the observed cluster abundance (Kitayama \&
Suto 1997).  We use the fitting formulae of Peacock \& Dodds (1996)
and Mo, Jing, \& B\"{o}rner (1997) for the nonlinear power spectrum
$P_{\rm R,nl}(k)$ and the peculiar velocity dispersions
$\sigmav$, respectively.

\begin{figure}[t]
\centerline{\epsfxsize=12cm \epsffile{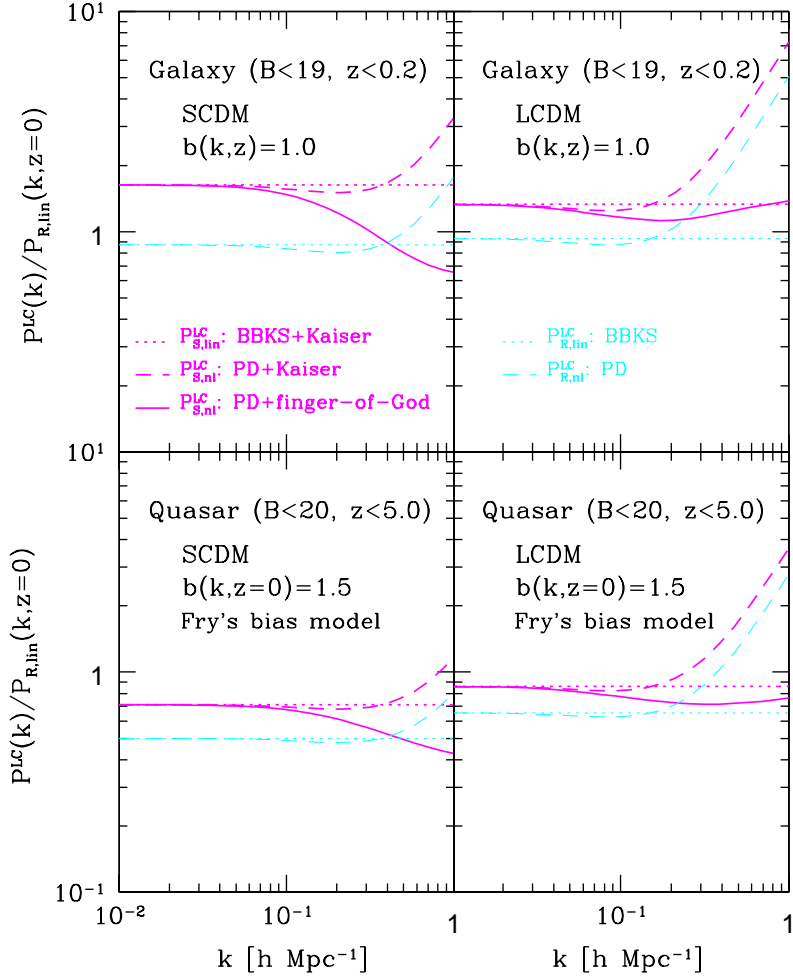}}
\caption{
    Power spectra on the light cone, $P^{\rm LC}(k)$, for
    various models, divided by $P_{\rm R,lin}(k,z=0)$, the
    corresponding linear power spectrum on the constant-time
    hypersurface in real space at $z=0$. Upper panels assume a galaxy
    sample with $B<19$ and $z<0.2$, while lower panels are for a
    quasar sample with $B<20$ and $z<5$, both of which roughly
    correspond to the upcoming SDSS spectroscopic samples. Left and
    right panels plot the results in SCDM and LCDM models,
    respectively. Thick and thin lines represent $P^{\rm LC}(k)$
    measured in redshift and real spaces.
}
    \label{fig:pklcGQ}
\end{figure}

Figure \ref{fig:pklcGQ} plots the theoretical predictions for power
spectra on the light cone. We adopt the B-band limiting magnitude $19$
and $20$ for galaxies and quasars, respectively, so as to roughly
match the SDSS and 2dF spectroscopic samples. For illustrative
purposes, the results are divided by the corresponding linear power
spectrum on the constant-time hypersurface in real space at $z=0$,
$P_{\rm R,lin}(k,z=0)$.  For simplicity we adopt a scale-independent
linear bias model of Fry (1996); $b(k,z=0)=1$ and $1.5$ for galaxies
and quasars ($p=1$ in eq.[\ref{FryM}] below).  Purely linear theory
predictions, $P^{\rm LC}_{\rm S,lin}(k)$, are plotted in dotted lines,
which use the CDM transfer function by Bardeen et al.(1986; BBKS) and
the linear redshift-space distortion (\ref{eq:PLCSlin}) by Kaiser
(1987).  We show two different versions of $P^{\rm LC}_{\rm S,nl}(k)$ ;
both adopt the Peacock \& Dodds (1996; PD) nonlinear spectrum, one,
plotted in dashed lines, uses linear redshift-space distortion, and
the other (solid lines) uses the nonlinear model (\ref{eq:PLCSnl})
with the velocity dispersion formula by Mo, Jing \& B\"{o}rner (1997).

In linear regime, the light-cone effect suppresses the amplitude of
spectrum in real space due to the smaller fluctuation at larger $z$
unless the possible evolution of bias exceeds the linear growth rate
of the mass fluctuations. The redshift-space distortion, on the other
hand, tends to increase the observable power spectrum in redshift
space. Thus the overall effect of the linear redshift-space distortion
on the light cone is to increase (decrease) the amplitude of $P^{\rm
  LC}_{\rm S,lin}(k)$ for galaxy (quasar) samples in a $k$-independent
manner.

The situation becomes complicated in nonlinear regimes, which produces
the additional $k$-dependent behavior. While the nonlinear evolution
of density field enhances the amplitude of both $P^{\rm LC}_{\rm R,nl}(k)$
and $P^{\rm LC}_{\rm S,nl}(k)$ relative to their counterparts in linear
theory , the suppression due to the finger-of-God is much stronger.
Consequently $P^{\rm LC}_{\rm S,nl}(k)$ is smaller than $P^{\rm LC}_{\rm
  S,lin}(k)$ for any $k$, and in fact the suppression is stronger for
larger $k$. Since recent numerical simulations (Mo, Jing, \&
B\"{o}rner 1997; Jing 1998; Suto et al. 1999; Magira et al. 1999)
confirmed that the nonlinear fitting formulae of density and velocity
fields are accurate and reliable within several percents, these
conclusions are generic as long as the bias is time-independent.  The
bias, however, is predicted to show strong time evolution. Furthermore
the realistic bias is expected to be neither time-independent nor
linear (Fry 1996; Mo \& White 1996); it may even not be deterministic
(Tegmark \& Peebles 1998; Dekel \& Lahav 1999; Taruya, Koyama \& Soda
1999). We will discuss this problem in detail elsewhere, and focus on
the effect of time-evolution of bias in the next subsection.

\subsection{Dependence on time-evolution of bias }

As discussed in previous subsection, the most uncertain aspect of the
theoretical predictions of $P^{\rm LC}(k)$ is the model of bias. Since
it is unlikely that any reliable theoretical model for bias is
established in near future, all that one can try is to explore the
possible effects by adopting a simple parametric model. While several
phenomenological models of evolution of bias have been proposed in the
literature (e.g., Moscardini, et~al. 1998; Matarrese, et~al. 1997), we
adopt the following simple model:
\begin{equation} 
  b(\eta)= 1 +{1\over [D_1(\eta)]^p} [b_0-1],
\label{FryM}
\end{equation}
with $b_0 \equiv b(\eta_0)$ is the bias parameter at present.  
When the constant index $p$ is unity, the model reproduces the
time-evolution of the lowest-order bias coefficient which is 
discussed by Fry (1996) on the basis of the continuity equation 
of mass and galaxy density fields.

\begin{figure}[t]
\centerline{\epsfxsize=12cm \epsffile{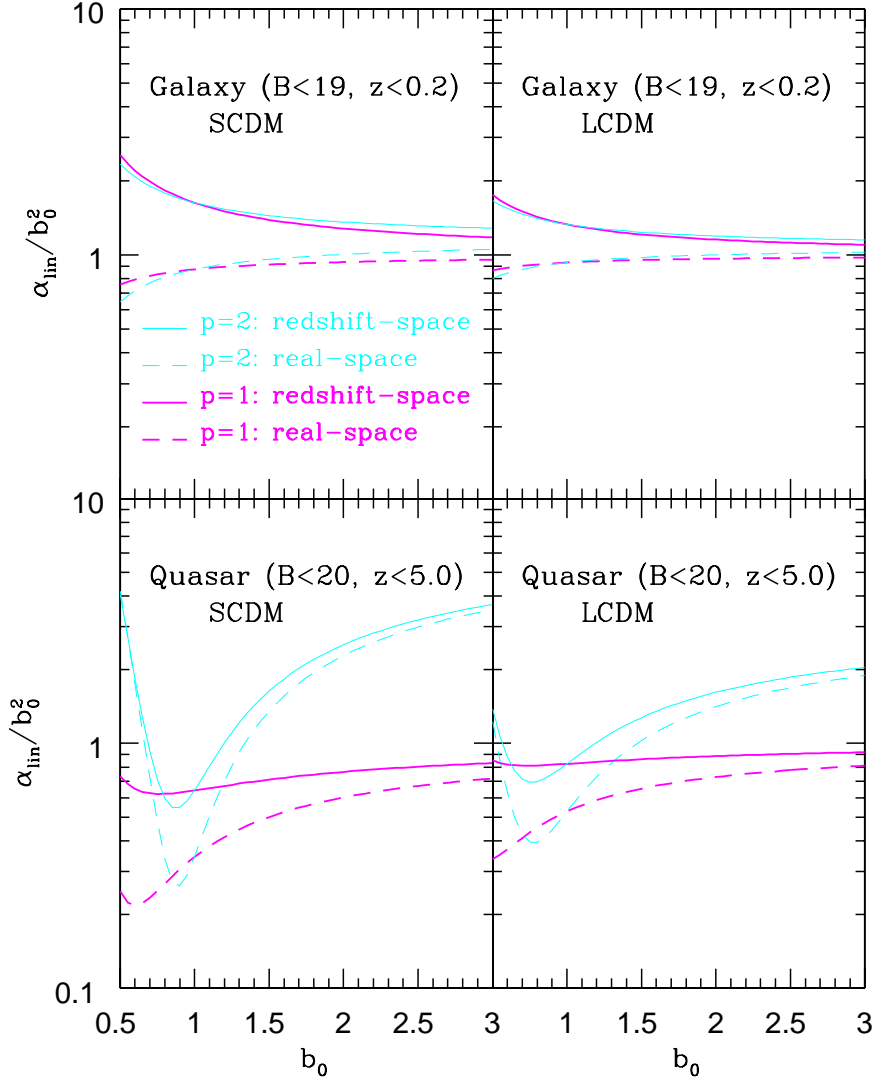}}
\caption{
    $\alpha_{\rm R,lin}/b_0^2$ and $\alpha_{\rm S,lin}/b_0^2$
    as a function of the linear bias parameter at present $b_0$.
    Solid and dashed lines represent the ratio measured in redshift
    and real spaces, respectively; $p=1$ in thick and $p=2$ in thin
    lines.  The model with $p=1$ corresponds to the bias evolution
    model of Fry (1996).
}
\label{fig:alphabias}
\end{figure}

Figure \ref{fig:alphabias} plots 
$\alpha_{\rm R,lin}/b_0^2$ and $\alpha_{\rm S,lin}/b_0^2$ for 
$p=1$ (Fry's model)
and $2$ (regarded as an extreme example for comparison) as a function
of $b_0$.  The selection functions for the galaxy and quasar samples
are identical to those adopted in Figure \ref{fig:pklcGQ}. For shallow
samples like the SDSS galaxies, theoretical predictions are fairly
insensitive to the applied model of evolution of bias; the
cosmological light-cone effect on these samples is dominated by the
redshift-space distortion.  For deeper samples, however, the amplitude
is very sensitive to the evolution of bias; especially when the bias
evolves faster than the linear growth rate ($p>1$), the redshift-space
distortion becomes relatively negligible on linear scales.

\section{Conclusions}

In this paper we have developed a theoretical formulation to compute
the power spectrum of cosmological objects on the light-cone
hypersurface. On the basis of the exact formula which works in linear
theory and in real space, we have obtained a useful approximate
expression valid on scales less than the survey volume size (Further
discussions for the validity of the approximation will be given
elsewhere; Nishioka \& Yamamoto 1999b). In linear theory, the
light-cone effect simply suppresses (in general) the amplitude of the
observable power spectrum (as in the case of the two-point correlation
functions; see Papers I, II and III).  Then we improved the
approximate formula by including the nonlinear evolution of density
fields, and linear and nonlinear redshift-space distortion
phenomenologically but in a manner fully consistent with the recent
numerical simulations. These nonlinearities produce the additional
scale-dependence in the power spectrum on the light cone compared with
those defined on the constant-time hypersurface.

Applying this expression to the galaxy and quasar samples selected so
as to match the upcoming SDSS spectroscopic samples, for instance, we
have quantitatively evaluated the degree of the light-cone effect.
With these example, we have demonstrated that we have developed a
fairly reliable and useful method to compute the power spectrum on the
light cone {\it given an evolution model of bias}.  In fact, it is
clear that the evolution of bias sensitively changes the behavior of
the power spectrum on the light cone, especially for quasar samples.
While our present analysis assumes the simplest bias model, linear
bias, we plan to extend our formulation for the general bias model
including the stochastic bias models (Dekel \& Lahav 1999; Taruya,
Koyama, \& Soda 1999)

\bigskip
\bigskip

We thank Y.Kojima, T.Matsubara and M.Sasaki for useful discussions and
comments, and Y.P.Jing and H.Magira for providing numerical routines
to compute fitting formulae for nonlinear density and velocity fields.
This research was supported in part by the Grants-in-Aid by the
Ministry of Education, Science, Sports and Culture of Japan to RESCEU
(07CE2002) and to K.Y. (11640280), by the Supercomputer Project
(No.98-35, and No.99-52) of High Energy Accelerator Research
Organization (KEK) in Japan, and by the Inamori Foundation.

\bigskip
\bigskip

\parskip2pt
\centerline{\bf REFERENCES}
\bigskip
\def\apjpap#1;#2;#3;#4; {\pp#1, {#2}, {#3}, #4}
\def\apjbook#1;#2;#3;#4; {\pp#1, {#2} (#3: #4)}
\def\apjppt#1;#2; {\pp#1, #2.}
\def\apjproc#1;#2;#3;#4;#5;#6; {\pp#1, {#2} #3, (#4: #5), #6}
\apjpap Bardeen,~J.M., Bond,~J.R, Kaiser,~N., \& Szalay,~A.S. 1986;
 ApJ;304;15 (BBKS);
\apjpap Cole, S., Fisher, K. B., \& Weinberg, D. H. 1995;MNRAS;275;515; 
\apjpap Davis, M. , \& Peebles, P.J.E. 1983;ApJ;267;465;
\apjpap Dekel, A. , \& Lahav, O. 1999;ApJ;520;24;
\apjpap de Laix, A. A. , \& Starkman, G. D. 1998;MNRAS;299;977;
\apjpap Efstathiou, G., Frenk, C.S., White, S.D.M.  , \& Davis, M.
 1988;MNRAS;235;715;
\apjpap Feldman, H. A., Kaiser, N., \& Peacock, A. A. 1994;ApJ;426;23;
\apjpap Fry, J. N. 1996;ApJ;461;L65;
\apjppt Hamilton, A.J.S. 1998; in `` The Evolving Universe. Selected
Topics on Large-Scale Structure and on the Properties of Galaxies'',
(Kluwer: Dordrecht), p.185;
\apjpap Jing, Y.P. 1998;ApJ;503;L9;
\apjpap Kaiser, N. 1987;MNRAS;227;1;
\apjpap Kitayama, T. , \& Suto, Y. 1997;ApJ;490;557;
\apjpap Loveday,J., Peterson, B.A., Efstathiou, G., \& Maddox, S.J. 
 1992;ApJ;390;338;
\apjppt Magira, H., Jing, Y.P., \& Suto, Y. 1999;ApJ, in press;
\apjpap Matarrese, S., Coles, P., Lucchin, F., \& Moscardini, L. 1997;
 MNRAS;286;115;
\apjpap Matsubara, T. , \& Suto, Y. 1996;ApJ;470;L1;
\apjpap Matsubara, T. , Suto, Y., \& Szapudi,I. 1997;ApJ;491;L1;
\apjpap Mo, H.J., Jing, Y.P. , \& B\"{o}rner, G. 1997;MNRAS;286;979;
\apjpap Mo, H.J. , \& White, S.D.M. 1996;MNRAS;282;347;
\apjpap Moessner, R., Jain, B., \& Villumsen, J. V. 1998;MNRAS;294;291;
\apjpap Moscardini, L., Coles, P., Lucchin, F., \& Matarrese, S. 1998
   ;MNRAS;299;95;
\apjpap Nakamura, T.T., Matsubara, T., \& Suto, Y. 1998;ApJ;494;13;
\apjpap Nakamura, T.T.,  \& Suto, Y. 1997;Prog. Theor. Phys.;97;49;
\apjpap Nishioka, H. , \& Yamamoto, K. 1999a;ApJ;520;426(Paper II);
\apjppt Nishioka, H. , \& Yamamoto, K. 1999b;in preparation;
\apjpap Peacock, J.A. , \& Dodds, S.J. 1994;MNRAS;267;1020;
\apjpap Peacock, J.A. , \& Dodds, S.J. 1996;MNRAS;280;L19 (PD);
\apjpap Suto, Y., Magira, H., Jing, Y. P., Matsubara, T., 
\& Yamamoto, K. 1999;Prog.Theor.Phys.Suppl.;
 133;183 (Paper III, astro-ph/9901179);
\apjpap Taruya, A., Koyama, K., \& Soda, J. 1999;ApJ;510;541;
\apjpap Tegmark, M. , \& Peebles, P.J.E. 1998;ApJ;500;L79;
\apjpap Ueda, H., Itoh, M., \& Suto, Y. 1993;ApJ;408;3;
\apjpap Wallington, S. , \& Narayan, R. 1993;ApJ;403;517;
\apjpap Yamamoto, K. , \& Suto, Y. 1999;ApJ;517;1 (Paper I);

\bigskip

\newpage
\parskip2pt
\centerline{\bf APPENDIX}
\begin{appendix}

\section{Linear theory derivation of the exact formula 
of the light-cone power spectrum $P^{\rm LC}_{\rm R,lin}(k)$ in real
space}

In this appendix, we outline the derivation of equation (\ref{PLC})
from equation (\ref{eq:defPkLC}):
\begin{eqnarray}
  P^{\rm LC}_{\rm R,lin}(k) 
={\displaystyle\int d^3{\bf r}_1 \int d^3 {\bf r}_2 
  \bigl<[n^{\rm LC}(r_1,\bft_1)-n_s^{\rm LC}(r_1,\bft_1)]
  [n^{\rm LC}(r_2,\bft_2)-n_s^{\rm LC}(r_2,\bft_2)]\bigr>
  e^{i {\bf k} \cdot ({\bf r}_1-{\bf r}_2)}
  \over \displaystyle\int d^3{\bf r} n_0^{\rm LC}(r)^2} .
\hspace*{0.4cm}
\label{eq:appdefPkLC}
\end{eqnarray}
With equations (\ref{nLC}) and (\ref{13}),the numerator of the
integrand in equation (\ref{eq:appdefPkLC}) becomes
\begin{eqnarray}
  &&\bigl< [n^{\rm LC}(r_1,\bft_1)-n_s^{\rm LC}(r_1,\bft_1)]
  [n^{\rm LC}(r_2,\bft_2)-n_s^{\rm LC}(r_2,\bft_2)]\bigr>
\nonumber
\\
 &&\hspace{3.5cm}=\n0LC(r_1) \n0LC (r_2)
  \bigl<\Delta(\eta_0-r_1,r_1,\bft_1)\Delta(\eta_0-r_2,r_2,\bft_2) \bigr>.
\label{3.5}
\end{eqnarray}
We expand the fluctuations $\Delta (\eta,r,\vec\gamma )$ 
in equation (\ref{defDelta}) as
\begin{equation}
  \Delta(\eta,\chi,\vec\gamma)=
  \int^{\infty}_0dk \sum_{l,m} \Delta_{klm}(\eta)
  {\cal Y}_{klm}(\chi,\bft),
\label{6}
\end{equation}
where ${\cal Y}_{klm}(\chi,\bft)$ is the normalized harmonics on the
flat space:
\begin{equation}
  {\cal Y}_{klm}(\chi,\bft)={\sqrt{2\over\pi}} 
  k j_l(k\chi)Y_{lm}(\Omega_{\vec\gamma}),
\end{equation}
with $j_l(x)$ being the spherical Bessel function and
$Y_{lm}(\Omega_{\vec\gamma})$ being the spherical harmonics on a unit
two-sphere. To proceed further, we adopt two assumptions; a linear
bias model in Fourier space:
\begin{equation}
  \Delta_{klm}(\eta)=b(k;\eta)\delta^{(c)}_{klm}(\eta) ,
\label{7}
\end{equation}
and linear growth of the density fluctuations:
\begin{equation}
\delta^{(c)}_{klm}(\eta)=\delta^{(c)}_{klm}(\eta_0)D_1(\eta),
\end{equation}
with $D_1(\eta)$ being the linear growth rate normalized to be
$D_1(\eta_0)=1$. In the above expressions, we denote the density
fluctuations by $\delta^{(c)}_{klm}(\eta)$ with CDM density
fluctuation specifically in mind, although not essential at all. 

Then the power spectrum of mass density fluctuations at present,
$P_{\rm R,lin}(k,\eta_0)$, is defined through
\begin{equation}
\bigl<\delta^{(c)}_{k_1l_1m_1}(\eta_0)\delta^{(c)*}_{k_2l_2m_2}(\eta_0)\bigr>
=P_{\rm R,lin}(k_1,\eta_0)\delta(k_1-k_2)\delta_{l_1l_2}\delta_{m_1m_2}.
\label{10}
\end{equation}
Substituting equations (\ref{6}) to (\ref{10}), one can write the
third term in equation (\ref{3.5}) as
\begin{eqnarray}
  &&\bigl<\Delta(\eta_0-r_1,r_1,\bft_1)\Delta(\eta_0-r_2,r_2,\bft_2)\bigr>
\nonumber
\\
  &&\hspace{2cm}={2\over\pi}\int^{\infty}_0dk_1k_1^2P_{\rm R,lin}(k_1,\eta_0)
  b(k_1;\eta_0-r_1)b(k_2;\eta_0-r_2)D_1(\eta_0-r_1)D_1(\eta_0-r_2)
\nonumber
\\
  &&\hspace{3cm}\times\sum_{l,m}j_l(k_1r_1)j_l(k_1r_2)
  Y^*_{lm}(\Omega_{\vec\gamma_1})Y_{lm}(\Omega_{\vec\gamma_2})~.
\label{DD}
\end{eqnarray}
Using the expansion of the plane wave in terms of the spherical harmonics:
\begin{equation}
e^{i{\bf k}\cdot {\bf r}}=4\pi\sum_{lm}i^lj_l(kr)
Y_{lm}(\Omega_{\vec\gamma})Y^*_{lm}(\Omega_{\hat{\bf k}}),
\label{15}
\end{equation}
and equation (\ref{DD}), equation (\ref{eq:appdefPkLC})
is explicitly written as
\begin{eqnarray}
P^{\rm LC}_{\rm R,lin}(k)&=&\left[\int d^3{\bf r} \n0LC(r)^2\right]^{-1}
\int dr_1 r_1^2 \int d\Omega_{\vec \gamma_{1}}
\int dr_2 r_1^2 \int d\Omega_{\vec \gamma_{2}}\n0LC(r_1)\n0LC(r_2)
\nonumber
\\
&~~& \times {2\over \pi} \int^{\infty}_0 dk_1k_1^2P_{\rm R,lin}(k_1,\eta_0)
b(k_1;\eta_0-r_1)b(k_1;\eta_0-r_2)D_1(\eta_0-r_1)D_1(\eta_0-r_1)
\nonumber
\\
&~~&\times \sum_{l,m}j_l(k_1r_1)j_l(k_1r_2)
Y^*_{lm}(\Omega_{\vec\gamma_1})Y_{lm}(\Omega_{\vec\gamma_2})
\nonumber
\\
&~~& \times ~ 4\pi\sum_{L_1M_1}i^{L_1}j_{L_1}(kr_1)
Y_{L_1M_1}(\Omega_{\vec\gamma_1})Y^*_{L_1M_1}(\Omega_{\hat{\bf k}})
\nonumber
\\
&~~& \times ~ 4\pi\sum_{L_2M_2}i^{-L_2}j_{L_2}(kr_2)
Y^*_{L_2M_2}(\Omega_{\vec\gamma_2})Y_{L_2M_2}(\Omega_{\hat{\bf k}}).
\label{16}
\end{eqnarray}

Integrating equation (\ref{16}) over $\Omega_{\vec\gamma_1}$ and
$\Omega_{\vec\gamma_2}$ results in
\begin{eqnarray} 
  &&P^{\rm LC}_{\rm R,lin}(k)=\left[\int d^3{\bf r} \n0LC(r)^2\right]^{-1}
  \int dr_1r_1^2 \int dr_2 r_2^2 \n0LC(r_1)\n0LC(r_2)
\nonumber
\\
  &&\hspace{1.5cm}\times{2\over \pi}\int^\infty_0dk_1k_1^2
  P_{\rm R,lin}(k_1,\eta_0)
  b(k_1;\eta_0-r_1)b(k_1;\eta_0-r_2)D_1(\eta_0-r_1)D_1(\eta_0-r_2)
\nonumber
\\
  &&\hspace{1.5cm}
  \times(4\pi)^2\sum_{l,m}j_l(k_1r_1)j_l(k_1r_2)j_l(kr_1)j_l(kr_2)
  Y_{lm}(\Omega_{\hat{\bf k}})Y^*_{lm}(\Omega_{\hat{\bf k}}).
\label{17}
\end{eqnarray}

Using the mathematical formula:
\begin{equation}
\sum^l_{m=-l}Y_{lm}(\Omega)Y_{lm}(\Omega ')
={2l+1\over 4 \pi}P_l(\cos \theta),
\label{18}
\end{equation}
one can simplify equation (\ref{17}) as follows:
\begin{eqnarray}
  &&P^{\rm LC}_{\rm R,lin}(k)= \left[\int dr r^2 \n0LC(r)^2\right]^{-1}
  \int dr_1r_1^2 \int dr_2 r_2^2 \n0LC(r_1)\n0LC(r_2)
  D_1(\eta_0-r_1)D_1(\eta_0-r_2)
  \nonumber
\\
  &&\hspace{1.5cm}\times{2\over \pi}\int^\infty_0dk_1k_1^2
  P_{\rm R,lin}(k_1,\eta_0)
  b(k_1;\eta_0-r_1)b(k_1;\eta_0-r_2)
  \nonumber
\\
  &&\hspace{1.5cm}\times
  \sum_l(2l+1)j_l(k_1r_1)j_l(kr_1)j_l(k_1r_2)j_l(kr_2).
\label{19}
\end{eqnarray}

Applying the formula:
\begin{equation}
  j_l(a \lambda) j_l(b \lambda)={1\over2\lambda}
  \int^1_{-1}dx (a^2+b^2-2abx)^{-{1 \over 2}} 
  \sin[\lambda(a^2+b^2-2abx)^{1\over 2}]P_l(x) 
\label{64}
\end{equation}
and then 
\begin{equation}
\delta(x-x')=\sum_n {2n+1 \over 2} P_n(x) P_n(x') ,
\label{65}
\end{equation}
equation (\ref{19}) is written as
\begin{eqnarray}
  P^{\rm LC}_{\rm R,lin}(k)&=&\left[\int dr r^2 \n0LC(r)^2\right]^{-1}
  \int dr_1r_1^2 \int dr_2 r_2^2 \n0LC(r_1)\n0LC(r_2)
  D_1(\eta_0-r_1)D_1(\eta_0-r_2)
  \nonumber
\\
  &&\hspace{.5cm}\times{2\over \pi}\int^\infty_0dk_1k_1^2
  P_{\rm R,lin}(k_1,\eta_0)
  b(k_1;\eta_0-r_1)b(k_1;\eta_0-r_2)  \sum_l(2l+1)
\nonumber
\\
  &&\hspace{.5cm}\times {1\over 2k_1}
  \int^1_{-1}dx_1(r_1^2+r_2^2-2r_1r_2x_1)^{-{1\over2}}
  \sin[k_1(r_1^2+r_2^2-2r_1r_2x_1)^{1\over2}]P_l(x_1)
\nonumber
\\
  &&\hspace{.5cm}\times {1\over 2k}
  \int^1_{-1}dx_2(r_1^2+r_2^2-2r_1r_2x_2)^{-{1\over2}}
  \sin[k(r_1^2+r_2^2-2r_1r_2x_2)^{1\over2}]P_l(x_2) 
\nonumber
\\
  &=& \left[\int dr r^2 \n0LC(r)^2\right]^{-1}
  \int dr_1r_1^2 \int dr_2 r_2^2 \n0LC(r_1)\n0LC(r_2)
  D_1(\eta_0-r_1)D_1(\eta_0-r_2)
\nonumber
\\
  &&\hspace{.5cm}\times{1\over \pi k}\int^\infty_0dk_1k_1 
  P_{\rm R,lin}(k_1,\eta_0)
  b(k_1;\eta_0-r_1)b(k_1;\eta_0-r_2)
\nonumber
\\
  &&\hspace{.5cm}\times\int^1_{-1}dx{\sin[k_1(r_1^2+r_2^2-2r_1r_2x)^{1/2}]
\sin[k(r_1^2+r_2^2-2r_1r_2x)^{1/2}]\over r_1^2+r_2^2-2r_1r_2x} .
\label{70}
\end{eqnarray}

Finally changing the variable $x$ to $y \equiv
(r_1^2+r_2^2-2r_1r_2x)^{1/2}$, equation (\ref{70}) reduces to
\begin{eqnarray}
  &&P^{\rm LC}_{\rm R,lin}(k)=\left[\int dr r^2 \n0LC(r)^2\right]^{-1}
  \int dr_1r_1 \int dr_2 r_2 \n0LC(r_1)\n0LC(r_2)
  D_1(\eta_0-r_1)D_1(\eta_0-r_2)
\nonumber
\\
  &&\hspace{0.5cm}\times{1\over \pi k}\int^\infty_0dk_1k_1 
  P_{\rm R,lin}(k_1,\eta_0)
  b(k_1;\eta_0-r_1)b(k_1;\eta_0-r_2)
  \int^{r_1+r_2}_{|r_1-r_2|}dy
  {\sin k_1y\sin ky \over y} .
\label{72}
\end{eqnarray}
Rewriting the last integral of the above expression in terms of the
the cosine integral function:
\begin{equation}
  C_i(x) \equiv -\int^\infty_x dt{\cos t\over t}, 
\end{equation} 
equation (\ref{72}) yields the final expression (\ref{PLC}), namely,
\begin{eqnarray}
  &&P^{\rm LC}_{\rm R,lin}(k)=\left[\int dr r^2 \n0LC(r)^2\right]^{-1}
  \int dr_1r_1 \int dr_2 r_2 \n0LC(r_1)\n0LC(r_2)
  D_1(\eta_0-r_1)D_1(\eta_0-r_2)
\nonumber\\
  &&\hspace{1.5cm}\times{1\over 2\pi k}\int^\infty_0dk_1k_1 
  P_{\rm R,lin}(k_1,\eta_0)
  b(k_1;\eta_0-r_1)b(k_1;\eta_0-r_2) \Pi(r_1,r_2,k_1,k) ,
\label{PLCA}
\end{eqnarray}
where 
\begin{eqnarray}
  &&                \Pi(r_1,r_2,k_1,k)=
                         -C_i\bigl(|r_1-r_2||k_1-k|\bigr)
                         +C_i\bigl(|r_1-r_2|(k_1+k)\bigr)
\nonumber\\
  &&\hspace{3cm}         +C_i\bigl((r_1+r_2)|k_1-k|\bigr)
                         -C_i\bigl((r_1+r_2)(k_1+k)\bigr)~.
\label{defPiA}
\end{eqnarray}
\end{appendix}

\newpage 

\end{document}